\documentclass[pdflatex,sn-nature]{sn-jnl}

\usepackage{graphicx}%
\usepackage{multirow}%
\usepackage{amsmath,amssymb,amsfonts}%
\usepackage{amsthm}%
\usepackage{mathrsfs}%
\usepackage[title]{appendix}%
\usepackage{xcolor}%
\usepackage{textcomp}%
\usepackage{booktabs}%
\usepackage{listings}%

\theoremstyle{thmstyleone}%

\theoremstyle{thmstyletwo}%

\theoremstyle{thmstylethree}%

\begin{document}

\title[Optical Spiking Neural Networks via Rogue-Wave Statistics]{Optical Spiking Neural Networks via Rogue-Wave Statistics}

\author[ ]{\fnm{Bahad{\i}r} \sur{Kesgin}}

\author[ ]{\fnm{G\"{u}ls\"{u}m Yaren} \sur{Durdu}}

\author*[ ]{\fnm{U\u{g}ur} \sur{Te\u{g}in}}\email{utegin@ku.edu.tr}

\affil[ ]{\orgdiv{Department of Electrical and Electronics Engineering}, \orgname{Ko\c{c} University}, \orgaddress{\city{\.{I}stanbul}, \postcode{34450}, \country{T\"{u}rkiye}}}

\abstract{Optical computing could reduce the energy cost of artificial intelligence by leveraging the parallelism and propagation speed of light. However, implementing nonlinear activation, essential for machine learning, remains challenging in low-power optical systems dominated by linear wave physics. Here, we introduce an optical spiking neural network that uses optical rogue-wave statistics as a programmable firing mechanism. By establishing a homomorphism between free-space diffraction and neuronal integration, we demonstrate that phase-engineered caustics enable robust, passive thresholding: sparse spatial spikes emerge when the local intensity exceeds a significant-intensity rogue-wave criterion. Using a physics-informed digital twin, we optimize granular phase masks to deterministically concentrate energy into targeted detector regions, enabling end-to-end co-design of the optical transformation and a lightweight electronic readout. We experimentally validate the approach on BreastMNIST and Olivetti Faces, achieving accuracies of 82.45\% and 95.00\%, respectively, competitive with standard digital baselines. These results demonstrate that extreme-wave phenomena, often treated as deleterious fluctuations, can be harnessed as structural nonlinearity for scalable, energy-efficient neuromorphic photonic inference.}

\maketitle

\section{Introduction}\label{intro}
The demand for high-speed, energy-efficient processing in artificial intelligence has driven a resurgence of interest in optical computing, which leverages the intrinsic parallelism and high bandwidth of light \cite{McMahon2023-ak, Wetzstein2020-pd}. This shift is motivated by the growing energy and policy considerations surrounding modern deep learning research \cite{strubell_energy_2020}. While early foundations were established in coherent optical processing and Fourier optics \cite{lugt1974coherent, goodman2005introduction, mcaulay1991optical}, recent advances in deep learning have catalyzed the development of novel architectures, most notably Diffractive Deep Neural Networks ($D^2NN$) \cite{Lin2018-sq}. These all-optical frameworks have been extended to fiber-based designs \cite{kesgin_fiber-based_2025}, associative memories based on the Hopfield model \cite{Farhat1985-zr}, and programmable random neural networks \cite{carpinlioglu_genetically_2025}. A critical challenge in optical computing remains the implementation of nonlinear activation functions, which are essential for deep learning but challenging to achieve at low optical powers. To address this, recent approaches have explored structural nonlinearity through multiple scattering \cite{Yildirim2024-cv, Xia2024-gf}, spatiotemporal mixing in multimode fibers \cite{Tegin2021-ao, Kesgin2025-sx}, and coherent nanophotonic circuits \cite{shen_deep_2017}. Additionally, frameworks such as Extreme Learning Machines (ELM) and reservoir computing have been successfully adapted to the optical domain to bypass complex training requirements \cite{huang2006extreme, pierangeli2021photonic, rafayelyan2020large}, pushing the boundaries of what linear optical systems can computationally achieve.

In parallel, optical rogue waves---rare, extreme-intensity events characterized by long-tailed statistics---have become a central theme in nonlinear and complex wave dynamics, with strong links to hydrodynamic analogies and universal mechanisms of intermittency \cite{akhmediev_roadmap_2016,dudley_instabilities_2014,dudley_rogue_2019}. Following early observations of heavy-tailed optical fluctuations \cite{solli_optical_2007}, a broad body of work has mapped how extreme events arise from modulation instability, breather-like dynamics, and spatiotemporal complexity across platforms ranging from multiple filamentation \cite{birkholz_spatiotemporal_2013} to fiber lasers \cite{liu_rogue_2015,tegin_real-time_2023}. Importantly, rogue-wave statistics can also have linear origins via caustics and wavefront-induced focusing, where interference and granularity seed intense localized peaks \cite{mathis_caustics_2015,arecchi_granularity_2011}, and nonlinear instability can further amplify these extremes \cite{safari_generation_2017}. Recent efforts have advanced quantitative characterization of intensity statistics \cite{racz_quantitative_2024} and demonstrated that multimode interactions can be engineered to tailor extreme-event formation in integrated settings \cite{durdu_engineering_2025}. Beyond their foundational interest, these controllable extremes suggest a route to implement threshold-like operations by converting rare high-intensity events into discrete computational primitives \cite{akhmediev_roadmap_2016,dudley_instabilities_2014}.

Simultaneously, Spiking Neural Networks (SNNs) have emerged as a promising neuromorphic paradigm, mimicking the event-driven, sparse processing of biological brains \cite{gerstner_spiking_2002}. SNNs offer superior energy efficiency and robustness compared to traditional artificial neural networks \cite{zhou_direct_2024, ding_neuromorphic_2025}, with training often facilitated by surrogate gradient methods \cite{che_differentiable_2022}. The translation of SNNs into the optical domain has led to diverse hardware implementations, including ultrafast spiking laser neurons based on Vertical-Cavity Surface-Emitting Lasers (VCSELs) \cite{robertson_toward_2020, hejda_neuromorphic_2021} and architectures incorporating plasticity \cite{han_delay-weight_2021}. Integrated photonic approaches have utilized phase-change materials and graphene-on-silicon structures to create on-chip spiking neurons \cite{feldmann_all-optical_2019, jha_photonic_2022, zhang_photonic-electronic_2024}. Furthermore, free-space optical SNNs are being developed to exploit spatial parallelism for large-scale neuromorphic tasks \cite{ahmadi_free-space_2024}, aiming to bridge the gap between biological plausibility and photonic speed.

In this work, we introduce an optical spiking neural network in which synaptic integration is implemented by free-space diffraction and spike generation is realized through programmable rogue-wave (caustic) events. Specifically, we exploit a physical homomorphism between diffractive propagation and the temporal integration of biological neurons, and use a trainable phase mask as complex synaptic weights to control where and when extreme-intensity caustics emerge. We define firing using a rogue-wave statistical criterion based on the significant intensity of the speckle field, converting rare high-intensity events into spatial spikes. To make this physics trainable, we develop a differentiable digital-twin model of the propagation and thresholding pipeline, enabling end-to-end co-training of the optical phase mask and an electronic readout. 
We validate that rogue-wave dynamics persist under deterministic data modulation and demonstrate the resulting optical SNN on benchmark classification tasks with experimental inference using an SLM-based free-space setup, showing strong agreement between simulation and experiment. 

\section{Results}\label{results}
\begin{figure}
    \centering
    \includegraphics[width=0.8\linewidth]{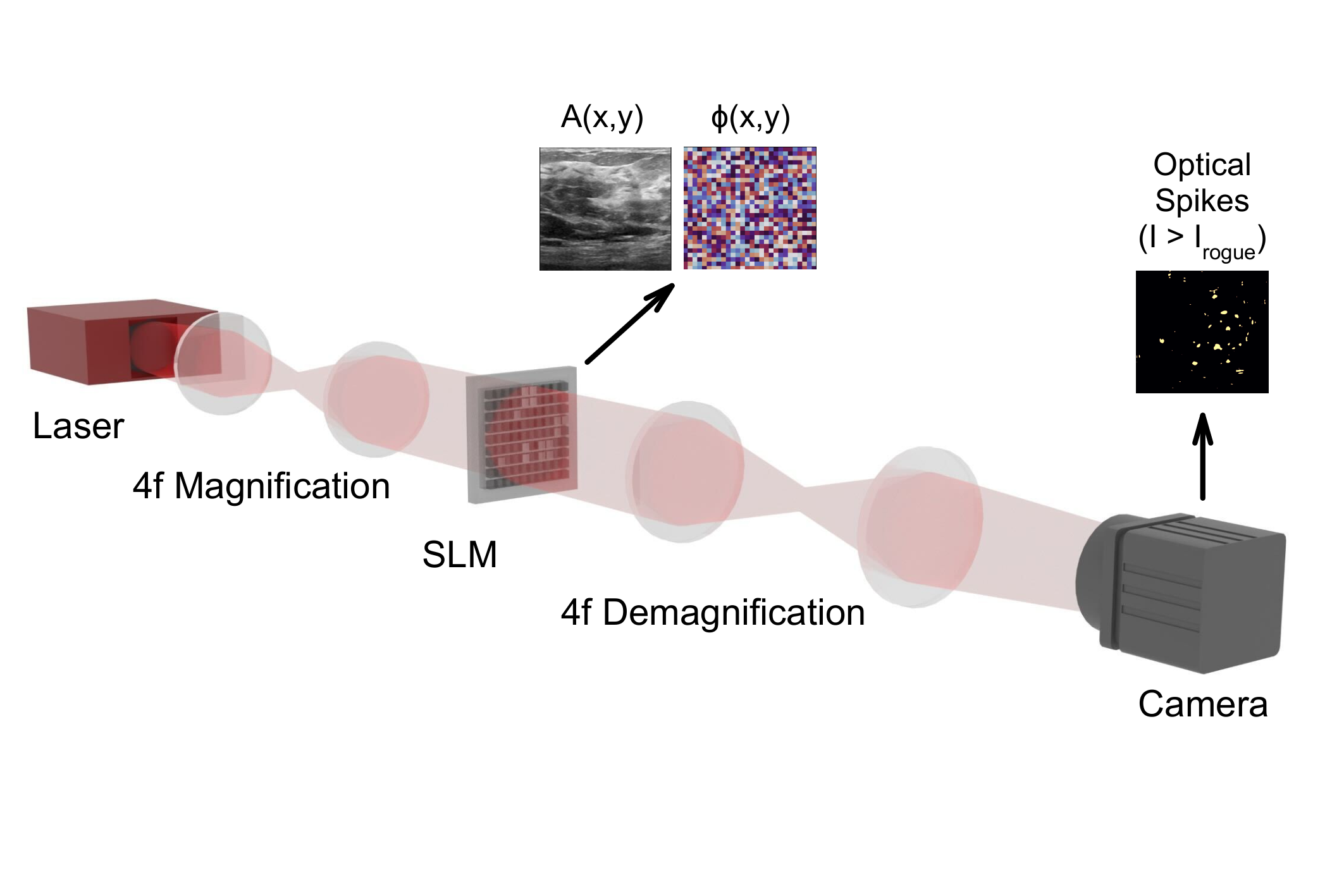}
    \caption{\textbf{Experimental schematic of rogue wave-based optical spiking neural network.} The collimated beam illuminates a reflective phase-only Spatial Light Modulator (SLM), which encodes the complex-valued input data and synaptic weights using a macropixel double-phase encoding scheme. A second calibrated 4-f relay system (L3-L4) demagnifies the diffracted speckle pattern to establish a 1-to-1 spatial correspondence between the SLM computation window and the CMOS detector array, ensuring accurate readout of the rogue wave events.}
    \label{fig:1}
\end{figure}
Experimental setup of our model is illustrated in Figure 1. In this experimental setup, the input information is encoded into the amplitude of a coherent optical field, while the synaptic weights are physically realized as a programmable phase mask on a Spatial Light Modulator (SLM). As the complex-modulated field propagates through free space, the passive physics of diffraction sums the secondary wavelets, leading to the emergence of spatial rogue waves—optical caustics formed by constructive interference. To translate this physical phenomenon into a robust computational architecture, we employ a co-training strategy that combines the optimization of the physical and digital parameters. The optical weights, phase of the pixels within the phase mask, are optimized via a physics-informed digital twin model to govern the formation of caustics, which are thresholded to generate spatial spikes which are processed in the co-trained readout layer.
\begin{figure}[htbp]
    \centering
    \includegraphics[width=\linewidth]{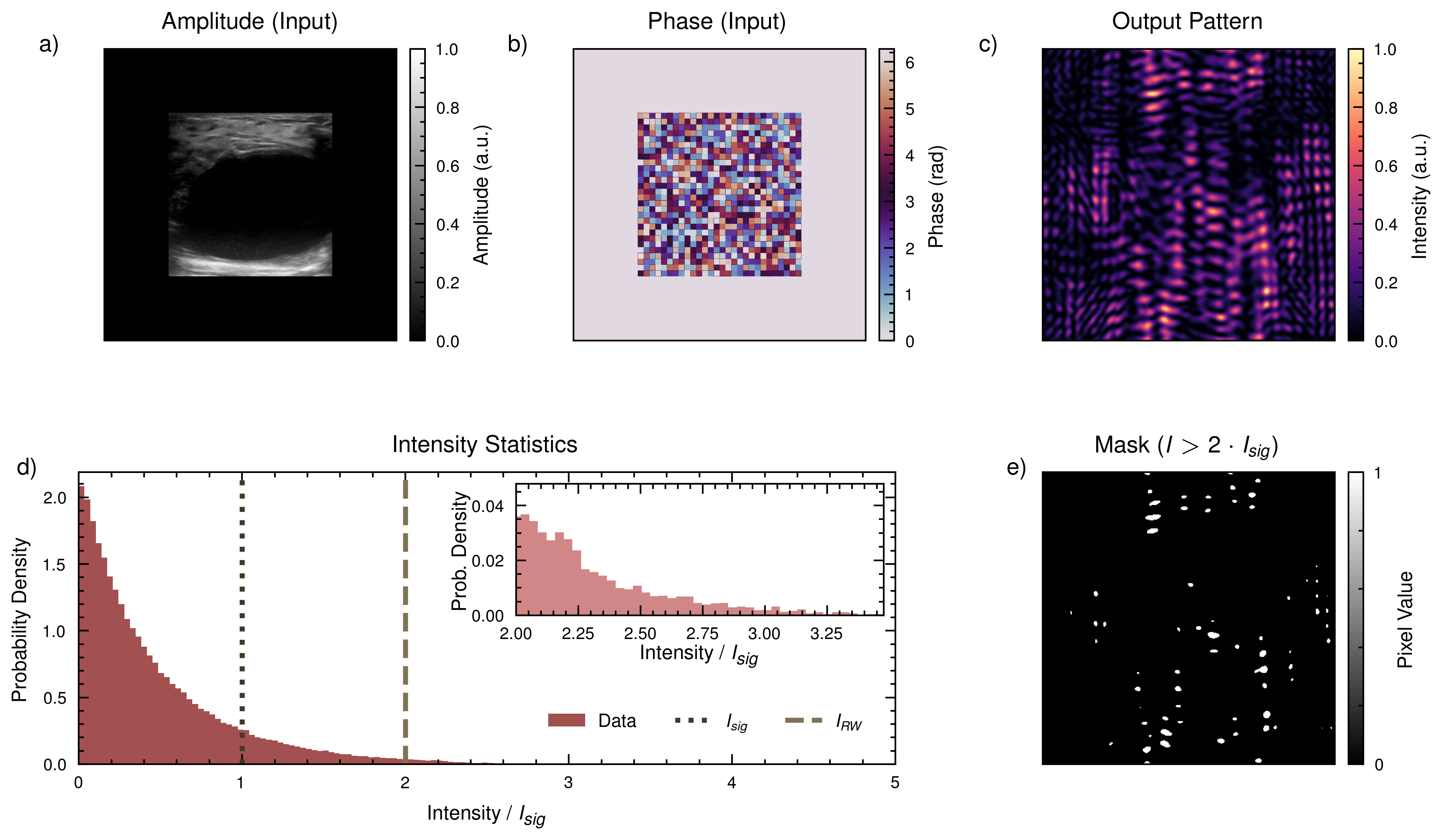}
    \caption{\textbf{Rogue waves in the presence of amplitude-encoded data}. \textbf{a}, Input amplitude distribution encoding the information. \textbf{b}, Phase modulation pattern applied to the SLM. \textbf{c}, Resulting optical intensity distribution at the detector plane after propagation. \textbf{d}, Histogram of the probability density function for the recorded intensity values, illustrating the statistical distribution.\textbf{e}, Optical spikes generated after rogue wave thresholding.}
    \label{fig:2}
\end{figure}
Prior investigations into optical rogue waves in free-space linear media typically rely on random phase modulation based initial wavefronts to generate long-tailed statistics; therefore, it is imperative to demonstrate that rogue wave dynamics persist and can be reliably controlled when the optical field is constrained by the deterministic amplitude distribution of real-world datasets. If the specific spatial frequencies inherent to the dataset encoded in the amplitude channel were to suppress the formation of caustics, the thresholding mechanism would fail, rendering the model inoperable. To verify this, we first conduct numerical simulations utilizing the BreastMNIST dataset. We first test 1000 random phase patterns to demonstrate rogue waves in our experimental configuration (see Supplementary Discussion 1 for details). Then we select one of these  granular phase patterns and perform simulations to test for rogue waves for data encoded in amplitude and control pattern encoded in phase channels. We observe that even with the imposition of complex amplitude modulation for data encoding, the application of the trained phase masks successfully triggers high-intensity caustic events at targeted spatial coordinates. While number of rogue waves and their maximum intensity changes image by image we manage to observe a rogue wave for every sample in the dataset for a known rogue wave generator phase. Our analysis confirms that for the specific phase distributions utilized in our network, the system can consistently operate in the heavy-tailed regime, ensuring robust and programmable rogue wave generation for the target samples. Figure \ref{fig:2} displays the spatial and statistical properties of the intensity pattern with the least ratio of $I_{max}/I_{significant}$.
\begin{table}[htbp]
\centering
\caption{Comparison of classification accuracies for BreastMNIST and Olivetti Faces datasets obtained by the proposed optical SNN and conventional digital benchmarks.}
\begin{tabular}{@{}lccc@{}}
\toprule
\textbf{Model} & \textbf{$n_{\text{Trainable Parameters}}$} &\textbf{BreastMNIST Acc.} \\ \midrule
\textbf{Proposed Optical SNN} & $\approx 20,000$ & \textbf{82.45\%} \\ 
ResNet-18  &  $11,000,000$ & 83.30\%  \cite{medmnistv2}\\
ResNet-50 & $25,000,000$ &84.20\%\cite{medmnistv2}\\
LeNet-5 & $61,000$ &80.76\%\cite{Kesgin2025-sx}\\
\bottomrule
\end{tabular}
\label{tab:1}
\end{table}
\begin{figure}[htbp]
    \centering
    \includegraphics[width=\linewidth]{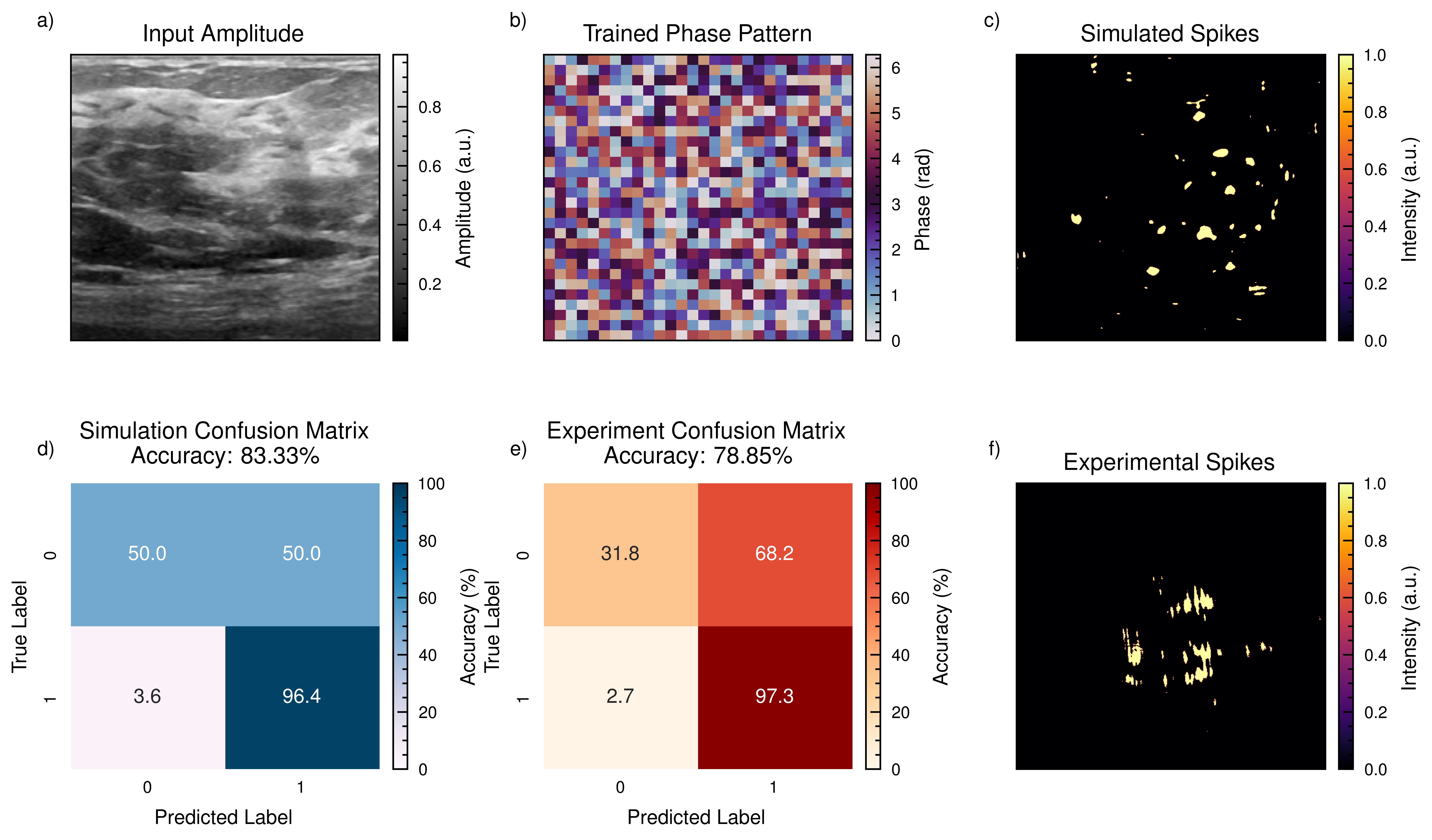}
    \caption{\textbf{Results of binary classification with BreastMNIST dataset.}textbf{a}, Input amplitude distribution encoding the information. \textbf{b}, Optimized phase modulation pattern applied to the SLM. \textbf{c}, Simulated optical spikes generated by the corresponding complex amplitude. \textbf{d}, Confusion matrix of the simulation results.\textbf{e}, Confusion matrix of the experimental results. \textbf{f}, Experimentally measured optical spikes generated by the corresponding complex amplitude.}
    \label{fig:3}
\end{figure}
Having established the physical validity of the caustic firing condition under data modulation, we evaluated the classification performance of the optical SNN using the BreastMNIST biomedical dataset. We deployed the phase masks, optimized via the digital twin, onto the SLM and performed inference using the optical setup, subsequently training the readout layer on the physically detected speckle patterns. The experimental results exhibited strong agreement with the numerical predictions, demonstrating the resilience of the rogue wave dynamics to experimental noise and optical aberrations. As illustrated in Table 1, our optical spiking neural network achieves competitive accuracy on the binary classification task, performing on par with conventional digital counterparts such as ResNet-18, ResNet-50 and LeNet-5. This result highlights the efficiency of the architecture, which performs the heavy-lifting of feature extraction and nonlinear thresholding entirely in the passive optical domain, leaving only a lightweight linear operation for the electronic backend.

\begin{figure}[htbp]
    \centering
    \includegraphics[width=\linewidth]{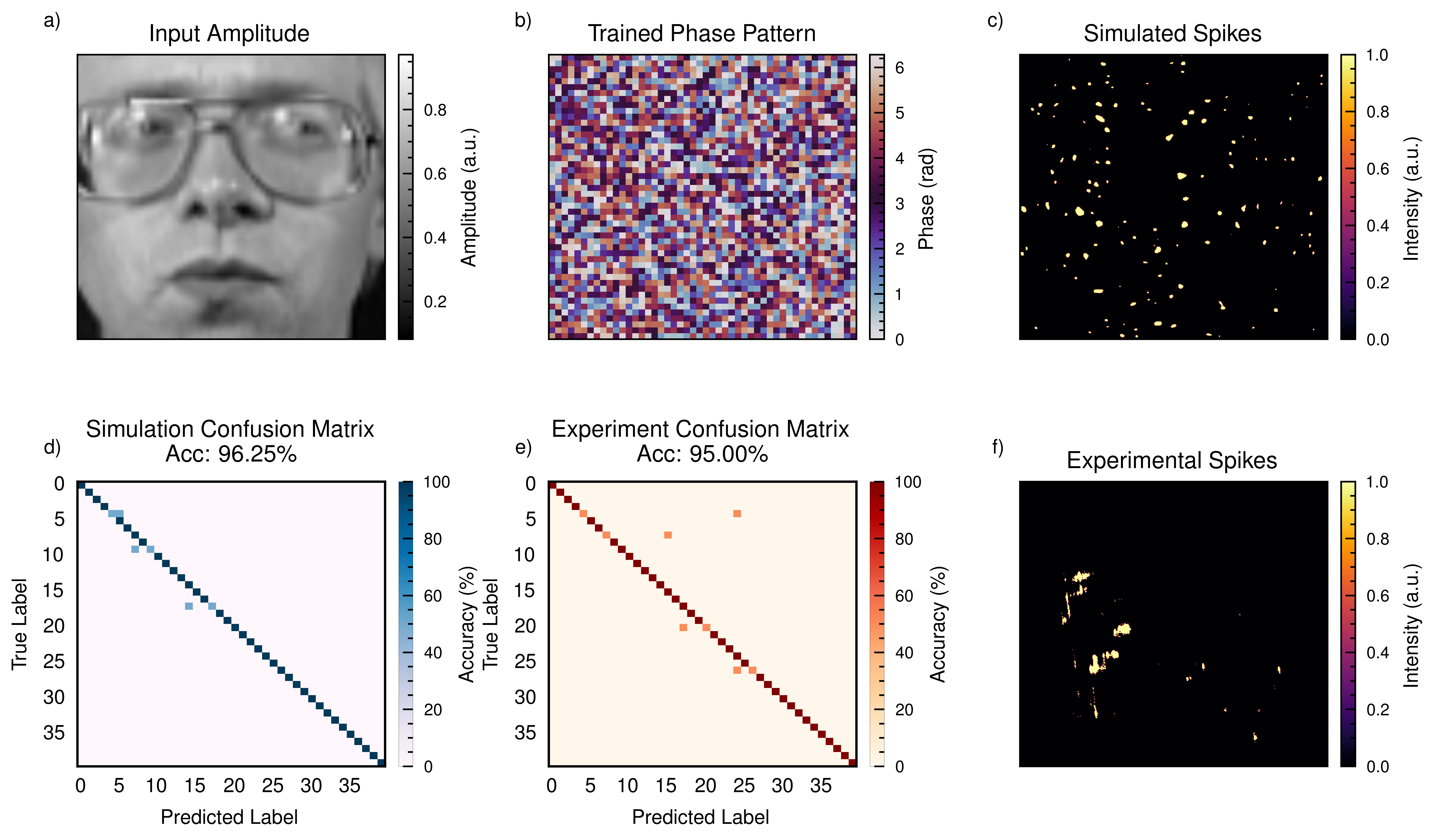}
    \caption{\textbf{Results of binary classification with Olivetti Faces dataset.}textbf{a}, Input amplitude distribution encoding the information. \textbf{b}, Optimized phase modulation pattern applied to the SLM. \textbf{c}, Simulated optical spikes generated by the corresponding complex amplitude. \textbf{d}, Confusion matrix of the simulation results.\textbf{e}, Confusion matrix of the experimental results. \textbf{f}, Experimentally measured optical spikes generated by the corresponding complex amplitude.}
    \label{fig:4}
\end{figure}
Encouraged by the successful implementation of the binary classification task, we extended our evaluation to a more challenging multi-class scenario using the Olivetti Faces dataset. This dataset requires the network to distinguish between distinct facial identities, demanding a higher degree of feature separability and precise control over the caustic formation to route energy to specific class-dependent detector regions. We trained the optical system to modulate the interference patterns such that the rogue wave spikes occur in spatially distinct zones corresponding to each identity. The results demonstrate that our model generalizes effectively to multi-class problems, achieving high classification accuracies on both the training and test sets (Table 1). This performance indicates that the diffractive optical reservoir can successfully map complex, high-dimensional input data into linearly separable rogue wave events, verifying the scalability of the proposed architecture for intricate computer vision tasks.

\section{Discussion}\label{discussion}
This work introduces an optical spiking neural network in which diffraction acts as synaptic integration and rogue-wave (caustic) statistics provide a physically grounded firing nonlinearity. By co-optimizing a phase-only diffractive mask with a differentiable digital twin, we program where rare, extreme-intensity events occur and convert them into sparse spatial spikes using the standard significant-intensity thresholding rule. The resulting optical front-end performs high-dimensional mixing and thresholding in a passive propagation stage, leaving only a lightweight electronic readout, while maintaining strong agreement between simulation and experimental inference across classification benchmarks.

Beyond the specific demonstrations, the central conceptual point is that \emph{extreme-event physics can be leveraged as a computational primitive}. Optical computing has long leveraged coherent propagation for fast linear transforms \cite{McMahon2023-ak,Wetzstein2020-pd,goodman2005introduction,Lin2018-sq}, but scalable nonlinear activation remains a key constraint; recent approaches show that effective nonlinearity can emerge from structured or recurrent linear optics when combined with detection and feedback \cite{Yildirim2024-cv,Xia2024-gf}. In parallel, rogue waves are a universal hallmark of complex wave systems, appearing across nonlinear and spatiotemporally rich optical platforms \cite{akhmediev_roadmap_2016,dudley_instabilities_2014,dudley_rogue_2019,solli_optical_2007,tegin_real-time_2023}. Our results connect these threads by using heavy-tailed caustic statistics to implement spike generation without requiring device-level excitability, complementing photonic SNN efforts based on spiking laser neurons and neurosynaptic photonic hardware \cite{robertson_toward_2020,feldmann_all-optical_2019,jha_photonic_2022}.

Looking forward, the principles established here are not limited to free-space optics. The homomorphism between spatial diffraction and temporal dispersion suggests that similar rogue-wave-based SNNs could be implemented in temporal domains using optical fibers or on-chip photonic circuits \cite{tegin_real-time_2023, Kesgin2025-sx}. Such implementations could leverage the high-bandwidth properties of integrated photonics to realize ultrafast neuromorphic processors \cite{robertson_toward_2020, zhang_photonic-electronic_2024}. By leveraging the rich physics of extreme wave phenomena, our work paves the way for a new class of "physically enhanced" computing architectures that harness the complexity of nature as a computational advantage.

\section{Methods}\label{methods}
\subsubsection*{Physical Model of the Optical Spiking Neuron}

The optical spiking neural network architecture we present harnesses the diffraction of light to execute synaptic integration passively, while exploiting the statistical properties of optical rogue waves as an efficient thresholding mechanism. We formulate the free-space propagation of the optical field as a linear transform governed by the Rayleigh-Sommerfeld diffraction integral. This framework treats propagation as the spatial summation of secondary spherical waves emitted from a source plane $(x', y')$ to a detector plane $(x, y)$ at a distance $z$.

We encode the input data into the spatial amplitude distribution $A_{in}(x',y')$, while the spiking control dynamics are governed by a programmable phase mask $\phi_{control}(x',y')$. This phase term functions as the complex synaptic weight, manipulating the interference patterns to trigger threshold events:
\begin{equation}
E_{in}(x', y') = A_{in}(x', y') \cdot e^{i\phi_{control}(x', y')}
\end{equation}

To demonstrate the similarities between optical diffraction and neuronal processing, we align the Rayleigh-Sommerfeld diffraction integral with the Spike Response Model (SRM)\cite{gerstner_spiking_2002}, a generalized integral formulation of the biological neuron. In the SRM, the membrane potential is defined relative to the last firing time $\hat{t}$, separating the integration of incoming postsynaptic potentials from the refractoriness induced by the previous spike. The spatial summation of the optical field is mathematically homologous to the temporal summation in the SRM:

\begin{equation}
E(x,y,z) = \iint^{\infty}_{- \infty } A_{in}(x',y')\cdot \omega_{c}(x',y') \cdot h_{opt}(x-x',y-y',z) \, dx' dy'
\end{equation}

\begin{equation}
u(t) = \eta(t-\hat{t}) + \int^{\infty}_{0} I(t-s) \cdot \omega_{s} \cdot \kappa(s) \, ds
\end{equation}

Here, the diffracted field $E(x,y,z)$ corresponds to the membrane potential $u(t)$, $\eta(t)$ defines the deterministic "reset" trajectory of the potential immediately following a spike, and the complex optical weight is given by $\omega_{c}(x',y') = e^{i\phi_{control}(x',y')}$. This complex optical weight is analogous to the synaptic efficacy $\omega_{s}$. These similarities reveal a structural homomorphism between the system kernels. The spherical wave propagator $h_{opt}$ corresponds to the neural response kernel $\kappa(s)$ (typically modeled as an exponential decay $e^{-s/\tau_m}$). Both functions describe how an input distributes and dissipates across the system's domain:
\begin{equation}
h_{opt}(x-x',y-y',z) = \frac{1}{i\lambda}\frac{ze^{ikr}}{r} \quad \longleftrightarrow \quad \kappa(s) = \frac{1}{C_m} e^{-s/\tau_m}
\end{equation}
Where $r=\sqrt{(x-x')^2 + (y-y')^2 + z^2}$. The firing mechanism of our optical neuron leverages the formation of optical caustics, also known as giant waves or rogue waves. These are statistically rare, high-intensity events emerging from the constructive interference of coupled modes or plane waves. We utilize the oceanographic definition of these extreme events to establish a physical firing threshold. First, we define the significant wave intensity, $I_{sig}$ (analogous to $H_s$ in hydrodynamics), as the mean intensity of the highest one-third of the optical speckle distribution:
\begin{equation}
I_{sig} \equiv I_{1/3} = \langle I(x,y) \rangle_{I \in \text{top } 33\%}
\end{equation}
Adhering to the standard criterion for rogue waves, we define the firing threshold $I_{RW}$ as twice the significant wave intensity ($I_{RW} = 2 \cdot I_{sig}$).
Consequently, a spike is generated only when the diffractive integration focuses energy into a caustic sufficiently intense to breach this statistical limit ($|E|^2 \ge I_{RW}$). 

\subsubsection*{Training of Optical SNN with Digital Twin}

To validate the proposed architecture and optimize the physical parameters of the optical system, we develop a physics-informed digital twin framework. To accurately model the diffraction of the optical field and create a differentiable model such that rogue wave masks are optimizable, we employed the Angular Spectrum Method (ASM)\cite{peng_neural_2020}, a spectral-domain technique that provides an exact solution to the Helmholtz equation for scalar fields. This method decomposes the complex input field into a superposition of plane waves via a two-dimensional Fast Fourier Transform (FFT), propagates each spectral component by applying a phase transfer function dependent on the spatial frequencies $(f_x, f_y)$, and reconstructs the diffracted field through an inverse FFT. We formulate this propagation operation as:
\begin{equation}
E_{prop}(x,y,z) = \mathscr{F}^{-1} \left\{ \mathscr{F} \left\{ E(x,y,0) \right\} \cdot e^{i 2\pi z \sqrt{\lambda^{-2} - f_x^2 - f_y^2}} \right\}
\end{equation}
Where $\mathscr{F}$ denotes the 2D Fourier Transform. Our \textit{in silico} model simulates the wave propagation, diffraction, and thresholding dynamics of the physical setup, allowing for end-to-end training via error backpropagation. The simulation models a physical computation window of $3.84 \times 3.84$ mm, discretized into a $480 \times 480$ grid with a pixel pitch of $8$ $\mu$m ($x_{res}, y_{res}$). We simulated the propagation of a coherent laser source with a wavelength of $\lambda = 635$ nm. The Gaussian beam profile was pre-calculated with a full-width at half-maximum (FWHM) of approximately $1.3$ mm to match the experimental source. The free-space optical propagation over a distance of $z=40$ cm was modeled using the Angular Spectrum Method (ASM), which provides an exact solution to the Helmholtz equation in the frequency domain.

We evaluated the generalization capability of the system using two distinct benchmarks, the FashionMNIST dataset and the BreastMNIST biomedical dataset, each requiring specific preprocessing to match the physical aperture. For the FashionMNIST task, we spatially aligned the low-resolution $28 \times 28$ inputs with the optical window by applying a $14\times$ upsampling operation, resulting in a final resolution of $384 \times 384$ pixels. Conversely, for the binary classification task using BreastMNIST, we utilized the medical images directly at their native resolution of $224 \times 224$ pixels to preserve the original diagnostic features without rescaling. All input images were subsequently zero-padded to the full $480 \times 480$ simulation window to prevent boundary artifacts during FFT-based propagation. The input data was encoded into the amplitude of the optical field, while the trainable network weights were represented by the phase mask. To induce granularity of phases for rogue wave generation, we implemented the phase mask using a checkerboard parameterization with a superpixel size of 8.

A critical component of the training was the implementation of a differentiable approximation of the rogue wave thresholding mechanism. While the physical definition of the firing condition involves a discrete inequality ($|E|^2 \ge I_{RW}$), such discontinuities prevent gradient flow. We formulated a soft gating mechanism where the threshold $I_{RW}$ is dynamically calculated for each sample as twice the mean intensity of the top $33\%$ of speckles. The spatial binary spike train was approximated by a steep sigmoid function:
\begin{equation}
I_{out} = \frac{1}{1 + \exp(-k(|E_{prop}|^2 - I_{RW}))}
\end{equation}
with a steepness factor $k=100.0$. This allowed the gradients to propagate through spike thresholding, and thus the optimization of the phase mask became feasible.

The optical output intensity ($I_{out}$ was processed by a digital readout layer consisting of an average pooling operation (kernel size $8 \times 8$) followed by a single linear classification layer. The model was trained end-to-end using the Adam optimizer with an initial learning rate of $1 \times 10^{-4}$ and a Cosine Annealing learning rate scheduler. We utilized the Cross-Entropy Loss function to penalize classification errors. The training was conducted for 200 epochs with a batch size of 20, employing an early stopping mechanism with a patience of 50 epochs to prevent overfitting. The entire pipeline was accelerated on a CUDA-enabled GPU.

\subsubsection*{Experimental Setup}

The experimental setup allows for the physical realization of the complex-valued spatiotemporal integration and rogue wave thresholding dynamics described in the physical model. The optical path is initiated by a continuous-wave laser diode (Thorlabs PL202) operating at a center wavelength of $\lambda = 635$ nm. To ensure a high-fidelity Gaussian spatial mode profile and eliminate high-frequency noise components, the beam is spatially filtered and expanded using a 4-f optical system equipped with a precision pinhole at the Fourier plane. The collimated beam subsequently illuminates a phase-only reflective Spatial Light Modulator (HOLOEYE Pluto 2.1 NIR-145). The SLM features a resolution of $1920 \times 1080$ pixels with an $8$ $\mu$m pixel pitch and 8-bit depth.

Although the SLM is a phase-only device, our network architecture requires the modulation of a complex input field $E_{in} = A_{in}e^{i\phi_{weight}}$. To achieve this full complex amplitude modulation, we employ the macropixel-based double phase encoding technique. Each logical unit of the optical field is synthesized by combining two adjacent phase pixels, where the resulting amplitude and phase are controlled by the relative phase difference and mean phase of the pair, respectively.

Following modulation, the optical field propagates through a free-space diffraction distance of $z=40$ cm, acting as the passive synaptic integration layer. To capture the resulting intensity distribution, we utilize a CMOS image sensor positioned at the detector plane. To ensure a precise spatial mapping between the diffractive weights and the detected speckle patterns, we implemented a second 4-f imaging system between the propagation volume and the camera. This post-processing relay is calibrated to demagnify the optical field, establishing a 1-to-1 spatial correspondence such that the $480 \times 480$ pixel computation window on the SLM is mapped directly to a matching $480 \times 480$ pixel region on the detector.

To effectively bridge the gap between numerical simulation and physical implementation, we use the  trained phase masks of the digital twin on the experimental validations. Due to inevitable experimental imperfections such as optical aberrations, SLM surface periodicity, and minor misalignment the physical transfer function deviates slightly from the ideal ASM propagation model. To mitigate these discrepancies, we trained the electronic readout layer using experimental data with the same data pipeline as the simulations but in the experimental regime.

\subsubsection*{Data availability}
Simulation data can be reproduced using the scripts at this Zenodo repository \cite{kesgin_utegin-lptopticalsnnwrw_2025}. Any additional data generated during experiments may be obtained from the authors upon reasonable request. Datasets used in the study are publicly available through their respective access links.

\subsubsection*{Code availability}
Codes related to the results in this work publicly are available at Zenodo \cite{kesgin_utegin-lptopticalsnnwrw_2025}.

\subsubsection*{Acknowledgments}
This work is supported by the Scientific and Technological Research Council of Turkey (T\"{U}B\.{I}TAK) under grant number 122C150.

\subsubsection*{Author contributions}
B.U.K. and G.Y.D. constructed the experimental setup and conducted numerical simulations and experiments. U.T. provided conceptual advice, and was in charge of supervision. All authors participated in the analysis of results and the writing of the manuscript.

\subsubsection*{Competing interests}
All other authors declare no competing interests.

\bibliography{references}

\end{document}